\documentclass[pra,twocolumn,showpacs,preprintnumbers,amsmath,amssymb,superscriptaddress, eqsecnum]{revtex4}
\usepackage{bm,graphicx,amsbsy,amsmath,amsfonts,amsthm,color,mathrsfs}

\newcommand{\beq}{\begin{equation}}
\newcommand{\eeq}{\end{equation}}
\newcommand{\ba}{\begin{eqnarray}}
\newcommand{\ea}{\end{eqnarray}}
\newcommand{\bea}{\begin{eqnarray}}
\newcommand{\eea}{\end{eqnarray}}
\newcommand{\bma}{\begin{subequations}}
\newcommand{\ema}{\end{subequations}}
\newcommand{\bwt}{\begin{widetext}}
\newcommand{\ewt}{\end{widetext}}

\begin{document}
%\preprint{APS/123-QED}

\title{Solid state multi-ensemble quantum computer in waveguide circuit model}

\author{Sergey A. Moiseev}
\email{samoi@yandex.ru}

\affiliation{Kazan Physical-Technical Institute of the Russian Academy of Sciences,
10/7 Sibirsky Trakt, Kazan, 420029, Russia}
\affiliation{Institute for Informatics of Tatarstan Academy of Sciences, 20 Mushtary, Kazan, 420012, Russia}

\affiliation{Physical Department of Kazan State University, Kremlevskaya 18, Kazan, 420008, Russia}

\author{Sergey N. Andrianov}
\affiliation{Institute for Informatics of Tatarstan Academy of Sciences, 20 Mushtary, Kazan, 420012, Russia}

\affiliation{Physical Department of Kazan State University, Kremlevskaya 18, Kazan, 420008, Russia}

\author{Firdus F. Gubaidullin}

\affiliation{Kazan Physical-Technical Institute of the Russian Academy of Sciences,
10/7 Sibirsky Trakt, Kazan, 420029, Russia}

\affiliation{Institute for Informatics of Tatarstan Academy of Sciences, 20 Mushtary, Kazan, 420012, Russia}

\date{\today}

%--------------------------------------------------------------------------------------------
% Abstract
%--------------------------------------------------------------------------------------------

\begin{abstract}

The first realization of solid state quantum computer was demonstrated recently
by using artificial atoms -- transmons in superconducting resonator. Here,
we propose a novel architecture of flexible and scalable quantum computer
based on a waveguide circuit coupling many quantum nodes of controlled atomic
ensembles. For the first time, we found the optimal practically attainable
parameters of the atoms and circuit for 100{\%} efficiency of quantum memory
for multi qubit photon fields and confirmed experimentally the predicted
perfect storage. Then we revealed self modes for reversible transfer of
qubits between the quantum memory node and arbitrary other nodes. We found a
realization of iSWAP gate via direct coupling of two arbitrary nodes with a
processing rate accelerated proportionally to number of atoms in the node. A
large number of the two-qubit gates can be simultaneously realized in the
circuit for implementation of parallel quantum processing. Dynamic coherent
elimination procedure of excess quantum state and collective blockade
mechanism are proposed for realization of $iSWAP$ and $\sqrt {iSWAP} $
quantum gates.

\end{abstract}
%for multi-mode quantum memories.


\date{\today}

\maketitle

\section{Introduction}
\label{sec:introduction}

Construction of large quantum computer (QC) is a synergetic physical and
engineering problem which imposes a number of critical requirements on
physical and spatial organization of interconnections between the qubits of
QC and with its near environment necessary for a quantum transmission and
readout of quantum calculations results \cite{Nielsen2000,Kaye2007}. Quantum computing is based
on delicate exploitation of number of various single- and two- qubit gates.
Usually, single qubit gates are relatively easily fulfilled experimentally
by using a well-known coherent control of single two-level atoms (atomic
qubit) or molecular qubits in external resonant electromagnetic fields \cite{Chuang1998,Jones1998,Chuang1998b}. Also, the single photon qubit gates can be realized by the linear
optics technique which provides simple procedures for rotation of the light
phase and polarization with a control of light by mirrors and
beamsplitters \cite {Knill2001,Kok2007}.

Most complicated problem is an experimental realization of a high enough
coupling constant between two arbitrary qubits in order to implement
the deterministic sufficiently fast two-qubit gates. Few promising approaches
have been proposed to increase a coupling constant of two qubit
interactions. For example, using a single mode optical cavity provides the
enhanced coupling constant of the interaction between single atom and
photon in the cavity \cite{Purcell1946,Berman1994}. Then cavity mediated photon-atom interaction that is also enhanced can be organized for implementation of basic quantum information processes \cite{Duan2004, Aoki2006}.Very powerful method to
increase the coupling constant with a photon is to use a Josephson qubits
characterized by superconducting current of mesoscopic magnitude \cite {Makhlin2001} and
by using the Josephson qubits (transmons) in superconducting
resonators \cite{Wallraff2004, Majer2007}. The two qubit transmon processor has been demonstrated
recently \cite{DiCarlo2009} for successful implementation of the Grover search and
Deutsch--Jozsa quantum algorithms.

Another promising tool of the coupling constant enhancement is an encoding
of qubits on multi-atomic coherent states. Here, the coupling constant of N
atoms with a photon can be enhanced by factor $\sqrt N $. Initially the
multi-atomic coherent ensembles have been used for optical quantum memory
(QM) \cite{Lukin2003,Julsgaard2004,Julsgaard2004a, Fleischhauer2005, Appel2008, Moiseev2001, Tittel2010, Lvovsky2009, Hammerer2010}, implementation of robust quantum communication over long
lossy channels \cite{Duan2001} and for single photon generation \cite{Kuzmich2003}. Recently
collective ensembles of multilevel systems have been discussed for QM
cooperated with quantum computing \cite{Rabl2006,Brion2007} . The idea has been developed to
use the multi-qubit QM integrated in a hybrid superconducting QC for
encoding of the qubits cooperated with transmon cooper pair box used for
quantum processing \cite{Wesenberg2009}. Here, the single and two qubit gates are
performed by coherent control of transmon qubit in the external
electromagnetic field and via two-qubit $iSWAP$ gate with the cavity photon
state. Complete quantum computing for large number of qubits is realized by
a transfer of any qubit pair on the transmon and photon qubits with their
subsequent swapping and using single qubit gates. The hybrid approach had
been proposed for quantum processing of more than 100 qubits stored in
atomic ensemble embedded in superconducting transmission line resonator.
Now, a spatial architecture of the multi-qubit transmon based on
superconducting QC is under consideration \cite{Helmer2009}.

\section{Principle scheme of multi-ensemble quantum computer in waveguide circuit
model}
\label{sec:Scheme}
Instead of the promising progress, the transmons are characterized by
relatively fast decoherent processes and can work only in superconducting
systems i.e. at very low temperatures that motivates a further search of new
principle schemes for realization of solid state QC. Here, we propose a
novel architecture of QC based on multi-atomic ensembles situated in
different spatial nodes in single mode QED cavity. We demonstrate the QC
architecture by using a waveguide circuit realization of the QED cavity which
yields possibility to use available technologies for realization of
multi-qubit QM and fast quantum processing satisfying the di-Vincenzo
criteria \cite{DiVincenzo1999}. In particular, scalability is easily achievable for the
proposed circuit architecture.

\begin{figure}
  \includegraphics[width=0.4\textwidth,height=0.3\textwidth]{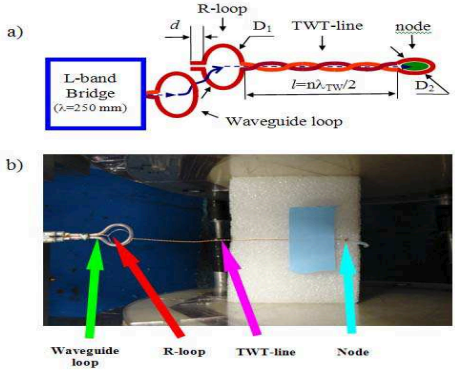}
  \caption{Scheme of experiment with waveguide circuit. a) L-band
Bridge is a source of microwave radiation coupled with primitive single node
circuit. The circuit consists of capacity of the R-loop tuned experimentally
by varying of length $d$, two waveguide transmission (TWT) line and small loop
surrounding the resonant medium (green spot) - quantum memory (QM) node;
D$_{1}$ and D$_{2}$ are the diameters ($D_2 / D_1 < < 1$). The radiation is
transferred from the L-band Bridge to the - (QM) node along arrow. b)
Realization of efficient field transfer from Waveguide loop via R-loop to
spatially distant Node for sufficiently large TWT-line length (rose);
waveguide loop, R-loop and node are pointed by green, red and blue arrows. }
  \label{Figure1}
\end{figure}

The primitive waveguide circuit is shown in Fig. \ref{Figure1}. The circuit consists of three
basic elements: a usual cupper receiver (R-) loop with diameter D$_{1}$,
two-waveguide transmission (TWT-) line and second loop with highly reduced
spatial diameter $D_{2}$ containing the resonant atomic ensemble (node). Two
waveguides of the TWT-line are twisted in order to suppress an irradiation to
external space. In our experiments, we used diameter $D_{2 }=0.4$ mm which
is more than 500 times smaller than the used wavelength $\lambda = 250$ mm
of the radiation transmitted through the external waveguide. The R-loop
provides an efficient reception of the radiation from the closely situated
waveguide output loop as shown in \ref{Figure1}. At small length of TWT-line, the
waveguide circuit works as a single mode resonator. We have been urged
experimentally in a robust work of the waveguide resonator in its numerous
spatial architectures. R-loop with inductivity L and capacity C tuned by
varying length of line $d$ (see\ref{Figure1}) determines a resonant frequency$\omega
_o = 1 / \sqrt {LC} $ of the circuit. TWT-line was adjusted to R-loop by
choosing its spatial length $\ell = n\lambda _{TW} / 2 = n\pi c / (\omega _o
\sqrt {\varepsilon _o \varepsilon } )$, where $\lambda _{TW} $ is a
wavelength of radiation in the TWT-line, c is the speed of light,
$\varepsilon $ is a nondimensional permittivity of TWT-line volume, $n$ is
an integer. We note that a distance between two twisted waveguides in TWT-line
was much smaller than the wavelength $\lambda _{TW} $ so the coupling
between TWT-line and R-loop didn't change the resonant frequency $\omega _o
$ while the line provided an effective transmission of the electromagnetic
field between the R- and small loops. Since the small loop had a negligibly
small spatial size the electromagnetic field evolved as a standing wave in
the TWT-line with the same electrical current in the circuit waveguide. Thus the
waveguide circuit works as a spatially distributed single mode resonator on the
resonant frequency $\omega _o $ with Q-factor $Q = r^{ - 1}\sqrt {L / C} $,
where r is a total losses resistance in the circuit (the losses were
negligibly weak in our experiments) and the circuit provides an effective
quantum electrodynamics of the atomic ensemble with electromagnetic field of
the single resonator mode.

\section{Multi-mode quantum memory}
\label{sec:MMQMemory}
By following cavity mode formalism \cite{Walls1994} we describe an efficient
multi-qubit QM in our circuit. We use the generalized Tavis-Cumming
Hamiltonian \cite{Tavis1968, Moiseev2007} $\hat {H} = \hat {H}_o + \hat {H}_1 $ for N atoms,
field modes and their interactions generalized by taking into account
inhomogeneous broadening of atomic frequencies and continuous spectral
distribution of the field modes where $\hat {H}_o = \hbar \omega _o
\{\sum\nolimits_{j = 1}^{N} {S_z^j } + \hat {a}_\sigma ^ + \hat {a}_\sigma +
\sum\nolimits_{n = 1}^2 {\int {\hat {b}_n^ + (\omega )\hat {b}_n (\omega
)d\omega } } \}$ are main energies of atoms ($S_z^j $ is a z-projection of
the spin operator), energy of cavity $\sigma $-field mode ($\hat {a}_\sigma
^ + $ and $\hat {a}_\sigma $ are arising and decreasing operators), energy
of waveguide field (n=1) and energy of free space field (n=2) ($b_n^ + $ and
$b_n $ are arising and decreasing operators of the waveguide modes),

\begin{align}
\label{eq1}
\hat {H}_1 = & \hbar \sum\limits_{j = 1}^N {\Delta _j S_z^j } + \hbar
\sum\limits_{n = 1}^2 {\int {(\omega - \omega _o )\hat {b}_n^ + (\omega
)\hat {b}_n (\omega )d\omega } }
\nonumber	\\	&
 + i\hbar \sum\limits_{j = 1}^N {[g_\sigma S_ - ^j \hat {a}_\sigma ^ + - }
g_\sigma ^\ast S_ + ^j \hat {a}_\sigma ^ - ]
\nonumber	\\	&
+ i\hbar \sum\limits_{n =
1}^2 {\int {\kappa _n (\omega )[\hat {b}_n (\omega )\hat {a}_\sigma ^ + -
\hat {b}_n^ + (\omega )a_\sigma ]d\omega } } .
\end{align}

The first two terms in (\ref{eq1}) comprise perturbation energies of atoms (where
$\Delta _j $ is a frequency detuning of j-th atom) and the field modes; the
third and fourth terms are the interaction energy of atoms with cavity mode
($S_ + ^j $ and $S_ - ^j $ are the transition spin operators, $g_\sigma $is
a coupling constant) and interaction energy of the cavity mode with the
waveguide and free propagating modes characterized by coupling constants
$\kappa _n (\omega )$.

We note that $[\hat {H}_o ,\hat {H}_1 ] = 0$ and Hamiltonian $\hat
{H}_o $ characterizes a total number of excitations in the atomic system and
in the fields which is preserved during the quantum evolution where $\hat
{H}_o $gives a contribution only to the evolution of common phase of
the wave function. H$_{1}$ determines a unitary operator $\hat {U}_1 (t) =
\exp \{ - i\hat {H}_1 t / \hbar \}$ causing a coherent evolution of the
atomic and field systems with dynamical exchange and entanglement of the
excitations between them. In spite of huge complexity of the compound
light-atoms system here we show that their quantum dynamics governed by
$\hat H_{1}$ in (1) can be perfectly reversed in time on our demand in a simple
robust way.

\begin{figure}
  \includegraphics[width=0.4\textwidth,height=0.3\textwidth]{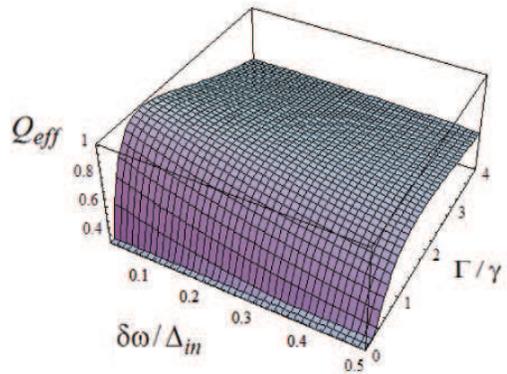}
  \caption{Transfer efficiency of
the input light field to the QM node from external waveguide as a function
of spectral width $\delta \omega / \Delta _{in} $ in units of inhomogeneous
broadening and ratio of $\Gamma / \gamma $ (where $\gamma_1+\gamma_2\approx\gamma_1 \equiv \gamma $) for Lorenzian spectral shapes of
the input field and inhomogeneous broadening for $\Delta _{in} =\gamma $.
It is seen that efficiency $Q_{eff} > 0.9$ for $\delta \omega / \Delta _{in}
< 0.2$ and the efficiency gets unity at $\Gamma / \gamma = 1$. }
  \label{Figure2}
\end{figure}

We assume that initially all atoms ($j = 1,2,...,N)$ stay on the ground
state $\left| 0 \right\rangle = \left| {0_1 ,0_2 ,...,0_N } \right\rangle $
and free mode fields are in the vacuum state and we launch a signal
multi-mode weak field to the circuit through the waveguide at t=0 as shown
in Fig.1. Total temporal duration of the signal field should be shorter than
decoherence time of the atomic system ($\delta t < < T_2 )$. We assume that
the spectral width $\delta \omega _f $ of the signal field is narrow in
comparison with inhomogeneously broadened width of the resonant atomic
transition ($\delta \omega _f < < \Delta _{in} )$. Then the consideration of
the atoms and field evolution for $\delta t < t < < T_2 $ gives an
efficiency $Q_{eff} $ of the signal field storage (a ratio of stored in
atomic system energy to incoming energy of the signal field) \cite{Moiseev2010a}:

\begin{equation}
\label{eq2}
Q_{eff} = \frac{\gamma _1 }{(\gamma _1 + \gamma _2 )}\frac{4\Gamma / (\gamma
_1 + \gamma _2 )]}{\vert 1 + \Gamma / (\gamma _1 + \gamma _2 )]\vert ^2},
\end{equation}

\noindent
plotted in Fig. \ref{Figure2}, where $\gamma _i = \pi \kappa _i^2 (\omega _o )$
determines a coupling between the resonator mode with waveguide modes (i=1)
and with free propagating modes (i=2) , $\Gamma = N_{qm} \vert g_\sigma
\vert ^2 / \Delta _{in} $- coupling of cavity mode with atomic system,
$N_{qm}$ is a number of atoms in the QM node.

As it is seen in Fig. \ref{Figure2}, the quantum efficiency $Q_{eff} $ reaches unity at
$\Gamma / \gamma _1 = 1$ and $\gamma _2 / \gamma _1 < < 1$ that shows a
promising possibility of perfect storage for multi-mode signal field at
moderate atomic density. Note that $\gamma _1 = \Gamma $ is a condition of
\textit{matched impedance condition }between the waveguide modes and the atomic system in QM node. Detailed
spectral analysis for the storage of the field characterized by finite
spectral width has been performed in paper \cite{Moiseev2010b}. Also, spectral matching
condition

\begin{equation}
\label{eq3}
\gamma _1 = 2\Delta _{in}
\end{equation}

\noindent
was found for the inhomogenenous broadening width $\Delta _{in} $ with
Lorentzian shape and the coupling constant $\gamma _1 $ that is a second
optimal condition for the QM. This condition provides a high efficiency of
the QM even in rather broad spectral range. In this case all multi-mode
signal fields incoming in the circuit transfer to the atomic system of the
QM node. The efficient direct unconditional transfer of the multi-mode field
is possible for inhomogeneously broadened atomic (electron spin) transition
where the effective quantum storage of multi-mode fields occurs for
arbitrary temporal profile of the modes.

We examined experimentally the signal storage for the radiation field with
carrier frequency $\nu = 1.2$ GHz. We varied parameter $\gamma _1 $ by
changing a spatial distance between R-loop and external waveguide loop in \ref{Figure1} and
found that all signal field energy was transferred to the electron spin
systems of lithium phthalocyaninate (LiPc) molecule sample at the optimal
matching value of $\gamma _1 = 3.768 \cdot 10^7$ where the reflected field
was absent in the external waveguide. At this condition we have observed a strong
signal of the electron paramagnetic resonance with a spectral width $\Delta
\nu = 88.2$ KHz from $4.51 \cdot 10^{13}$ LiPc molecules situated in one
distant node in Fig. \ref{Figure1}.

In order to construct an efficient QM for the multi-mode fields we follow
the original protocol of the photon echo QM proposed in 2001 \cite{Moiseev2001} and
theoretically described \cite{Moiseev2007} in most general way in the Schr\"{o}dinger
picture by exploiting symmetry properties of the light atoms Hamiltonian.
Here, we exploit simplicity of this approach in description of the
multi-mode QM in the proposed QC. The assumed atomic detunings $\Delta _j $
are caused by a presence of the magnetic field gradient. By assuming a
perfect storage in accordance with above coupling matching condition we
change a sign of the detunings $\Delta _j \to - \Delta _j $ at time moment
t=t' by changing of the magnetic field polarity similar to recent
experiments \cite{Hosseini2009}. By using a substitution for the field operators $\hat
{a}_\sigma = - \hat {A}$ and $\hat {b}_n (\omega _o - \Delta \omega ) = \hat
{B}_n (\omega _o + \Delta \omega )$ (with similar relations for the Hermit
conjugated operators) we get a new Hamiltonian $\hat {H}_1 ' = - \hat {H}_1
$ which has an opposite sign with respect to initial Hamiltonian determining
a reversed quantum evolution in accordance with a new unitary operator $\hat
{U}_2 [(t - t')] = \exp \{ - i\hat {H}_1 '(t - t') / \hbar \} =  \quad \exp
\{i\hat {H}_1 (t - t') / \hbar \}$. So the initial quantum state of the
multi-mode signal field will be reproduced at $t = 2t'$ in the echo pulse
with the irradiated field spectrum inverted relatively to the frequency
$\omega _o $ in comparison with the original one. Thus we have shown for the
first time that the multi-mode QM can be realized with nearly 100{\%}
efficiency by using the optimal matching condition $\gamma _1 = \Gamma $ of
the atoms in circuit with the waveguide modes. The possibility opens a door
for practical realization of ideal multi-mode QM.

\begin{figure}
  \includegraphics[width=0.4\textwidth,height=0.3\textwidth]{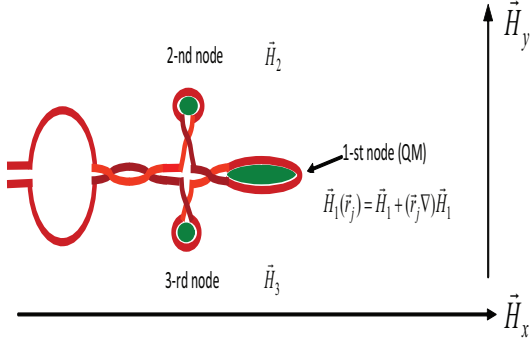}
  \caption{Principle scheme of three node quantum computer.
Left node is a QM node loaded in gradient magnetic field; Second and third
nodes are the processing nodes situated in the constant different magnetic fields.}
  \label{Figure3}
\end{figure}

Let's consider a principle scheme of the quantum transport in QC
architecture or three node circuit depicted in Fig. \ref{Figure3}.

The circuit contains
the QM node and two processing nodes. QM node is loaded in gradient magnetic
field providing an inhomogeneous broadening of atomic frequencies $\Delta
_{in} > > \delta \omega _f $ with central atomic frequency coinciding with
the circuit frequency $\omega _{qm} = \omega _o $. The second and third
nodes have N atoms in each node with equal frequencies $\omega _{2,3} $
within each node tuned far away from the frequency $\omega _o $. We should
provide a perfect reversible transfer of arbitrary qubits between QM node
and each other two nodes. Initially, the multi-qubit states encoded in the M
temporally separated photon modes $E(t) = \sum\nolimits_{m = 1}^M {E_m (t)}
$ with spectral width $\delta \omega _f $ are recorded in the QM-node by
excitation from the external waveguide. When the storage procedure is
completed we tune away the atomic frequency of the QM node from resonance
with the circuit $\omega _{qm} \ne \omega _o $. In order to transfer one
arbitrary k-th qubit state from the QM node to the second node we switch off
the waveguide circuit coupling with the external waveguide ($\gamma _1 = 0$ ) and
launch rephasing of the atomic coherence in QM node (by reversion of the
atomic detunings $\Delta \to - \Delta )$. The k-th qubit state will be
rephased at t=2t$_{k}$. At time moment t$_{k }$+t$_{k - 1}$ we equalize the
frequencies of QM-node and 2-nd node with $\omega _o $. The quantum dynamics
of atoms in QM and 2-nd nodes and circuit mode evolves to complete transfer
of atomic excitation from the QM node to the second-node at t=2t$_{k}$ when
the temporal shape of rephased single photon wave packet is

\begin{equation}
\label{eq4}
E_k (t) = E_o \exp \{ - \Gamma \vert t - 2t_k \vert / 2\}\sin S(t - 2t_k )/S,
\end{equation}

\noindent
where $S = \sqrt {N\vert g_\sigma \vert ^2 - (\Gamma / 2)^2}$, $t < 2t_k $, E$_{o}$ is an arbitrary small field amplitude.

\begin{figure}
  \includegraphics[width=0.4\textwidth,height=0.3\textwidth]{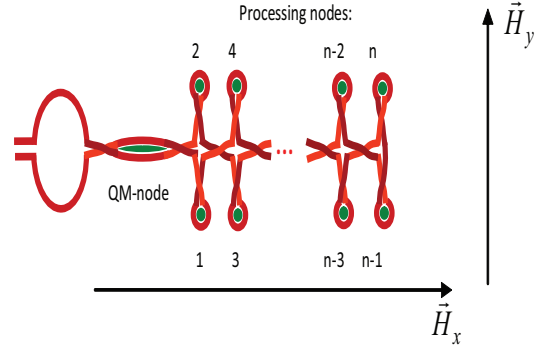}
  \caption{Multi-qubit quantum computer. Magnetic fields
$\vec {H}_x $ and $\vec {H}_y $ are used to control atomic frequencies in
the nodes; atoms in the QM node exist in the magnetic field gradient that
leads to inhomogeneous broadening of the atomic frequencies. The magnetic
field magnitude in the nodes is varied during the quantum processing for
rephasing the atomic coherence and to control the resonance conditions in
the circuit-node and node-node interactions.}
  \label{Figure4}
\end{figure}

These modes are the self quantum modes of the QC that provide the perfect
reversible coupling of multi-mode QM with processing nodes. Similar temporal
shape was proposed recently for single-mode QM \cite{Kalachev26}. After qubit transfer,
we switch off the coupling of the 2-nd node with resonator by changing the
frequency of atoms in the 2-nd node. The same procedure can be fulfilled for
transfer of qubit from QM node to the $3^{rd}$ node. The described picture
of the quantum transport can be realized in more complicated 2D scheme of QC
with larger number of nodes as depicted in Fig. \ref{Figure4}.

\section{Quantum processing}
\label{sec:MMQMemory}

For realization of two-qubit gates we transfer the two qubits from QM node
to the 2-nd and 3-rd processing nodes and equalize the carrier frequencies
of the nodes at time moment t=0 with some detuning from the resonator mode
frequency $\omega _2 - \omega _o = \Delta _2 = \omega _3 - \omega _o =
\Delta _3 = \Delta $. It yields to the following initial state of 2-nd and
3-rd nodes in the interaction picture

\begin{equation}
\label{eq5}
\psi _{in} (0) = \{\alpha _2 \left| 0 \right\rangle _2 + \beta _2 \left| 1
\right\rangle _2 \}\{\alpha _3 \left| 0 \right\rangle _3 + \beta _3 \left| 1
\right\rangle _3 \},
\end{equation}

\noindent
where $\vert \alpha _{2,3} \vert ^2 + \vert \beta _{2,3} \vert ^2 = 1$.
Here, we have introduced the following states: $\left| 0 \right\rangle _m =
\vert 0_1 ,0_2 ,...,0_{N_m } \rangle $ corresponding to the ground state of
the m-th node, $\left| 1 \right\rangle _m = \sqrt {1 / N} \sum\nolimits_{j=1}^{N_m
} {\left| {0_1 } \right\rangle \left| {0_2 } \right\rangle ...\left| 1
\right\rangle _j ...\left| {0_{N_m } } \right\rangle } $ and $\left| 2
\right\rangle _m = \sqrt {2 / N(N - 1)} \sum\nolimits_{i \ne j}^{N_m } {\left|
{0_1 } \right\rangle \left| {0_2 } \right\rangle ...\left| 1 \right\rangle
_i ...\left| 1 \right\rangle _j ...\left| {0_{N_m } } \right\rangle } $ are
the collective states of m-th node with single and two atomic excitations.
Equal frequencies of the two nodes results in interaction of the atoms via
the virtual processes of resonant circuit quanta determined by effective
Hamiltonian \cite{Imamoglu1999, Schuch2003} $\hat {H}_{eff} = \sum\nolimits_{m = 1}^3 {\hat
{H}_{node}^{(m)} } + \hat {H}_{int} $, where $\hat {H}_{node}^{(m)} = \hbar
\Omega _\sigma \sum\nolimits_{i_m, j_m }^N {S_{i_{1m} }^ + S_{j_m }^ - } $ is
a long-range spin-spin interaction in m-th node, $\hat {H}_{int} = \hbar
\Omega _\sigma \sum\nolimits_{j_1 ,j_2 = 1}^N {\left( {S_{j_1 }^ + S_{j_2 }^
- + S_{j_1 }^ - S_{j_2 }^ + } \right)} $ (where $\Omega _\sigma = \vert
g_\sigma \vert ^2 / \Delta )$ describes a spin-spin interaction between the
two nodes ($N_2 = N_3 = N)$. The spin-spin interaction with atoms of QM-node
is suppressed because of the absence of resonance.

Let's introduce the collective basis states of the two nodes:$\left| \psi
\right\rangle _1 = \left| 0 \right\rangle _2 \left| 0 \right\rangle _3 $,
$\left| \psi \right\rangle _2 = \left| 1 \right\rangle _2 \left| 0
\right\rangle _3 $, $\left| \psi \right\rangle _3 = \left| 0 \right\rangle
_2 \left| 1 \right\rangle _3 $, $\left| \psi \right\rangle _4 = \left| 1
\right\rangle _2 \left| 1 \right\rangle _3 $ and $\left| \psi \right\rangle
_5 = (1 / \sqrt 2 )\{\left| 2 \right\rangle _2 \left| 0 \right\rangle _3 +
\left| 0 \right\rangle _2 \left| 2 \right\rangle _3 \}$. It is important
that the Hamiltonian $\hat {H}_{eff} $ has a matrix representation in the
basis of the five states which is separated from other states of the
multi-atomic system

\begin{equation}
\label{eq6}
\left( {{\begin{array}{*{20}c}
 0 \hfill & 0 \hfill & 0 \hfill & 0 \hfill & 0 \hfill \\
 0 \hfill & {N\Omega _\sigma } \hfill & {N\Omega _\sigma } \hfill & 0 \hfill
& 0 \hfill \\
 0 \hfill & {N\Omega _\sigma } \hfill & {N\Omega _\sigma } \hfill & 0 \hfill
& 0 \hfill \\
 0 \hfill & 0 \hfill & 0 \hfill & {2N\Omega _\sigma } \hfill & {2\Omega
_\sigma \sqrt {N(N - 1)} } \hfill \\
 0 \hfill & 0 \hfill & 0 \hfill & {2\Omega _\sigma \sqrt {N(N - 1)} } \hfill
& {2\Omega _\sigma (N - 1)} \hfill \\
\end{array} }} \right).
\end{equation}

By using initial state (\ref{eq5}) we find the unitary evolution of the atomic systems which
couples independently two pairs of the quantum states $\left| \psi
\right\rangle _2 \leftrightarrow \left| \psi \right\rangle _3 $ and $\left|
\psi \right\rangle _4 \leftrightarrow \left| \psi \right\rangle _5 $

\begin{align}
\label{eq7}
 \Psi _1 \left( t \right) = &\alpha _2 \alpha _3 \left| \psi \right\rangle _1 +
 \exp ( -i\Omega _\sigma Nt)
\nonumber \\ &
\{\beta _2 \alpha _3 [\cos (\Omega _\sigma Nt)\left| \psi \right\rangle _2
- i\sin (\Omega _\sigma Nt)\psi _3 ]
\nonumber \\ &
 \mbox{ } + \alpha _2 \beta _3 [\cos (\Omega _\sigma Nt)\left| \psi \right\rangle _3
 - i\sin (\Omega _\sigma Nt)\psi _2 ]\}
\nonumber \\ &
 \mbox{ } + \exp ( - i2\Omega _\sigma Nt)\beta _2 \beta _3
\nonumber \\ &
\{\cos (2\Omega_\sigma Nt)\left| \psi \right\rangle _4
- i\sin (2\Omega _\sigma Nt)\left| \psi \right\rangle _5 \},
\end{align}

\noindent
where we have assumed a large number of atoms $N\gg 1$. The solution
demonstrates two coherent oscillations with the frequency $\Omega _\sigma N$
for the first pair $\left| \psi \right\rangle _2 \leftrightarrow \left| \psi
\right\rangle _3 $ and with the double frequency $2\Omega _\sigma N$ for the
second pair $\left| \psi \right\rangle _4 \leftrightarrow \left| \psi
\right\rangle _5 $. The oscillations are drastically accelerated N-times
comparing to the case of two coupled two-level atoms so we can use even bad
common resonator with relatively lower quality factor. By taking into
account our experimental situation and by assuming $\Delta \approx 10\Delta
_{in} $ we get $\Omega _\sigma N = \gamma _1 / 10 = 1.884 \cdot 10^6$.

It is known \cite{Imamoglu1999, Schuch2003} that the evolution of the two coupled two level atoms
can lead to $iSWAP$ and $\sqrt {iSWAP} $ gates. The $iSWAP$ and $\sqrt
{iSWAP} $ gates work in the Gilbert space of four states $\left| \psi
\right\rangle _1 ,...,\left| \psi \right\rangle _4 $ and these gates are
important for realization of the complete set of the universal quantum
gates \cite{Imamoglu1999, Schuch2003}. $iSWAP$ gate provides exchange of the two quantum states
between the two nodes. In our case we get iSWAP gate occurs at shortened
time $t_{iSWAP} = \pi / 2\Omega _\sigma N \cong 8.34 \cdot 10^{ - 7}$ sec.

\begin{equation}
\label{eq8}
\Psi _1 \left( {t_{iSWAP} } \right) = \{\alpha _3 \left| 0 \right\rangle _2
- \beta _3 \left| 1 \right\rangle _2 \}\{\alpha _2 \left| 0 \right\rangle _3
- \beta _2 \left| 1 \right\rangle _3 \}.
\end{equation}

We also note that by choosing different carrier frequencies we can realize
the described iSWAP operation for many pairs of nodes simultaneously due to
exploitation of the independent virtual quanta for each pair\underline { }in
the QED cavity. It is interesting that the $iSWAP$ gate provides a perfect
elimination of transfer of the initial state to the state $\left| \psi
\right\rangle _5 $ that occurs only at t=t$_{iSWAP}$. The situation is more
complicated for realization of $\sqrt {iSWAP} $ gate because it is
impossible to eliminate state $\left| \psi \right\rangle _5 $ with evolution
based on matrix (\ref{eq4}). Below, we propose a universal mechanism for \textit{Collective Dynamical Elimination}
(CDE --procedure) of the state $\left| \psi \right\rangle _5 $ for realization of
$\sqrt {iSWAP} $ gate by using the multi-atomic ensemble encoding single
qubits and cavity mediated collective interaction.

\begin{figure}
  \includegraphics[width=0.4\textwidth,height=0.3\textwidth]{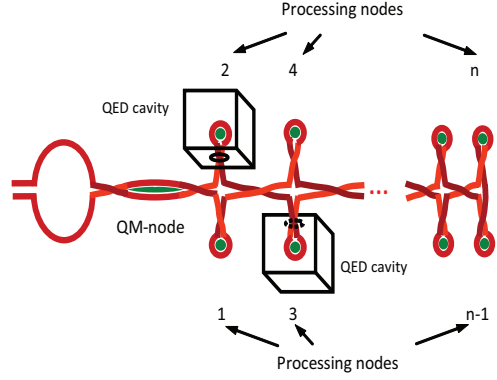}
  \caption{Scheme of QC with processing gates for realization
of $\sqrt {iSWAP} $\textbf{ gate. }Scheme of QC with two (2-nd and 3-rd)
processing nodes situated in local QED cavities characterized by the high
quality factors. QM and all processing nodes are coupled via the common
TW-line of the distributed electromagnetic resonator.}
  \label{Figure5}
\end{figure}

Scheme of spatial arrangement of the processing nodes and cavities for
realization of the $\sqrt {iSWAP} $ is presented in Fig. \ref{Figure5}.

 Here, we insert
the 2-nd and 3-rd nodes in two different single mode QED cavities
characterized by high quality factors for $\pi $-modes. We assume that each
$\pi $-mode interacts only with the atoms of one node and is decoupled from
the basic circuit field mode that is possible for large enough spectral
detuning of the local QED cavity modes. Thus we take the additional field
Hamiltonians
$H_{\pi } = \sum\nolimits_m^{2,3} {\hbar \omega _{\pi_m }
\hat a_{\pi _m }^ + \hat a_{\pi _m } } $
and the modes interaction with the atoms of
2-nd and 3$^{rd}$ nodes
$H_{r - a}^{\left( \pi \right)} =
\sum\nolimits_m^{2,3} {\sum\nolimits_{j_m=1 }^{N_m } {\left( {g_{\pi }^{\left( m
\right)} S_{j_m }^ + \hat a_{\pi _m } + g_{\pi}^{\left( m \right)\ast
} S_{j_m }^ - \hat  a_{ \pi _m }^ + } \right)} } $
%$H_{r - a}^{\left( \pi \right)} =
%\sum\limits_m^{2,3} {\sum\limits_{j_m } {\left( {g_{k_0 \pi _m }^{\left( m
%\right)} S_{j_m }^ + a_{k_0 \pi _m } + g_{k_0 \pi _m }^{\left( m \right)\ast
%} S_{j_m }^ - a_{k_0 \pi _m }^ + } \right)} } $ (where $g_{\pi }^{\left( m
%\right)}$
is a coupling constant of the interaction between atom and local m-th $\pi$ mode).
By assuming a large enough
spectral detuning of atomic frequencies from the field mode and absence of
real photons in the QED cavities we find the following effective Hamiltonian
similar to previous section

\begin{align}
\label{eq95}
H_s & = \sum\limits_m^{2,3} {\sum\limits_{j_m=1 }^{N_m } {\hbar \omega _m S_{j_m
}^z } }
\nonumber  \\  &
 + \sum\limits_m^{2,3} {\sum\limits_{i_m ,j_m }^{N_m } {\frac{\left|
{g_\sigma ^{\left( m \right)} } \right|^2}{\hbar \Delta _m }S_{i_m }^ +
S_{j_m }^ - } + \sum\limits_m^{2,3} {\sum\limits_{i_m ,j_m }^{N_m }
{\frac{\left| {g_\pi ^{\left( m \right)} } \right|^2}{\hbar \Delta _m^{\prime}
}S_{i_m }^ + S_{j_m }^ - } } }
\nonumber \\  &
 + \frac{1}{2\hbar }\left( {\frac{1}{\Delta _2 } + \frac{1}{\Delta _3 }}
\right)\sum\limits_{j_1 j_2 }^{N_m } {(|g_\sigma ^{\left( 2 \right)} g_\sigma
^{\left( 3 \right)}| e^{i\varphi (j_2 ,j_3 )}S_{j_2 }^ + S_{j_3 }^ - }
\nonumber  \\  &
 + g_\sigma ^{\left( 2 \right)\ast } g_\sigma ^{\left( 3 \right)} e^{ -
i\varphi (j_2 ,j_3 )}S_{j_2 }^ - S_{j_3 }^ + ),
\end{align}

\noindent
where $\Delta _{2,3} = \omega _{2,3} - \omega _o $ are the atomic frequency
detunings from the circuit mode and $\Delta _{2,3}^{\prime} = \omega _{2,3} -
\omega _{\pi} $  are the atomic detunings from the frequency of the local
QED cavities having the same frequency $\omega _{\pi} $ ($\omega _{\pi_2}=\omega _{\pi_3 } =\omega _{\pi}  $);
$\varphi (j_2 ,j_3 )$ is a phase which is assumed to be constant for the node size
smaller than the mode wavelength. To be concrete
below we take $\Delta _{2,3}^{\prime} = - \Delta _{2,3} = - \Delta $, $\Delta > 0$.

The second and third terms in Eq. (\ref{eq7}) describes the atom-atom interactions
insight each node via the exchange of $\sigma $ and $\pi $ virtual photons,
while the last term describes the interaction due to the exchange of virtual
$\sigma $ photons between the atoms seated in different nodes. Again by
assuming equal number of atoms in the two nodes $N_2 = N_3 = N$, we get the
following matrix representation for the new effective Hamiltonian $\hat
{H}_{eff} $ in the basis of the five states

\begin{equation}
\label{eq10}
\left( {{\begin{array}{*{20}c}
 0 \hfill & 0 \hfill & 0 \hfill & 0 \hfill & 0 \hfill \\
 0 \hfill & {\Omega _s N} \hfill & {\Omega _\sigma N} \hfill & 0 \hfill & 0
\hfill \\
 0 \hfill & {\Omega _\sigma N} \hfill & {\Omega _s N} \hfill & 0 \hfill & 0
\hfill \\
 0 \hfill & 0 \hfill & 0 \hfill & {2\Omega _s N} \hfill & {2\Omega _\sigma
\sqrt {N(N - 1)} } \hfill \\
 0 \hfill & 0 \hfill & 0 \hfill & {2\Omega _\sigma \sqrt {N(N - 1)} } \hfill
& {2\Omega _s (N - 1)} \hfill \\
\end{array} }} \right),
\end{equation}

\noindent
where $\Omega _s = \Omega _\sigma + \Omega _\pi $, $\Omega _\sigma = {\vert
g_\sigma \vert ^2} \mathord{\left/ {\vphantom {{\vert g_\sigma \vert ^2}
\Delta }} \right. \kern-\nulldelimiterspace} \Delta $, $\Omega _\pi = -
{\vert g_\pi \vert ^2} \mathord{\left/ {\vphantom {{\vert g_\pi \vert ^2}
\Delta }} \right. \kern-\nulldelimiterspace} \Delta $.

For the initial state (\ref{eq5}), the atomic wave function evolves as follows

\begin{align}
\label{eq11}
\Psi _2 (t) = &\alpha _2 \alpha _3 \left| \psi \right\rangle _1 +
\exp [ - i\Omega _sNt]
\nonumber \\  &
\{\beta _2 \alpha _3 [\cos (\Omega _\sigma Nt)\left| \psi \right\rangle _2
- i\sin (\Omega_\sigma Nt)\left| \psi \right\rangle _3 ]
\nonumber \\  &
\mbox{ } + \alpha _2 \beta _3 [\cos (\Omega _\sigma Nt)\left| \psi \right\rangle _3
- i\sin (\Omega _\sigma Nt)\left| \psi \right\rangle _2 ]\}
\nonumber \\  &
+ \exp [ - i\Omega _s (2N - 1)t]\beta _2 \beta _3
\nonumber \\  &
\{[\cos (St) - i\frac{\Omega _s }{S}\sin (St)]\left| \psi \right\rangle _4
\nonumber \\  &
- i\frac{2\Omega _\sigma \sqrt {N(N -1)} }{S}\sin (St)\left| \psi \right\rangle _5 \},
\end{align}

\noindent
where $\mbox{S} = \sqrt {4\Omega _\sigma ^2 N(N - 1) + \Omega _s^2 } $.

We choose the following parameters for the evolution of (\ref{eq11}) providing
the dynamical elimination of the state $\left| \psi \right\rangle _5  $

\begin{equation}
\label{cond1}
1)  \quad  \Omega _\sigma Nt = \pi (\textstyle{1 \over 4} + \textstyle{1 \over2}\mu + n); \mu = 0,1; n = 0,1,...,
\end{equation}

\begin{equation}
\label{cond2}
2)  \quad  St = \pi k, k = 1,2,...,
\end{equation}

\noindent
that leads to the following entangled state of the nodes

\begin{align}
\label{eq13}
\Psi _2 (t)& =  \alpha _2 \alpha _3 \left| \psi \right\rangle _1
\nonumber \\ &
+ ( - 1)^n\textstyle{1\over {\sqrt 2 }}\exp [ - i\Omega _s Nt]
\{[( - 1)^\mu \beta _2 \alpha _3 -
i\alpha _2 \beta _3 ]\left| \psi \right\rangle _2
\nonumber \\ &
+ [( - 1)^\mu \alpha _2 \beta _3 -
i\beta _2 \alpha _3 ]\left| \psi \right\rangle _3 \}
\nonumber \\ &
 + ( - 1)^k\exp [ - i\Omega _s (2N - 1)t]\beta _2 \beta _3 \left| \psi \right\rangle _4 ,
\end{align}

\noindent
where $\Omega _s $ is determined by two conditions \ref{cond1}, \ref{cond2}. In
particular we write three set of parameters for possible realizations of CDE
procedure characterized by weaker coupling of atoms with $\sigma $-mode
(n=0,1; $\mu $=0,1):

\begin{equation}
\label{eq14}
\begin{array}{l}
 1) n = 0, \mu = 0,k = 1: \Omega _{\sigma} N t =  \pi/ 4,
 S t = \pi \\
 \rightarrow |\Omega _s | t = \sqrt {3} \pi ,
 \frac{| \Omega _s | }{\Omega _{\sigma} N} = 4\sqrt {3} \approx 6.92; \\

 2) n = 0, \mu = 1,k = 2: \Omega _{\sigma} Nt = 3\pi / 4,
 St = 2\pi \\
 \rightarrow |\Omega _s | t = \sqrt {7} \pi,
 \frac{|\Omega _s |}{\Omega _{\sigma} N} = \frac{4\sqrt {7} }{3} \approx 5.53; \\

 3) n = 1, \mu = 0, k = 3: \Omega _{\sigma} Nt = 5\pi / 4,
 St = 3\pi \\
 \rightarrow | \Omega _s | t = \sqrt {11} \pi ,
 \frac{| \Omega _s |}{\Omega _{\sigma} N} =\frac{4\sqrt {11} }{5} \approx 2,65. \\
 \end{array}
\end{equation}

Another interesting case occurs for stronger coupling of the atoms with
local $\pi $-modes of the QED cavities when $\vert \Omega _\pi \vert > >
N\Omega _\sigma $. Here, we get a \textit{Collective Blockade} of state $\left| \psi \right\rangle _5$ that provides the
following atomic evolution

\begin{align}
\label{eq15}
 \Psi _2 (t) = & \alpha _2 \alpha _3 \left| \psi \right\rangle _1
 + \exp [ - i\Omega _sNt]
 \nonumber \\ &
 \{\beta _2 \alpha _3 [\cos (\Omega _\sigma Nt)\left| \psi \right\rangle _2 - i\sin (\Omega
_\sigma Nt)\left| \psi \right\rangle _3 ]
\nonumber \\ &
 \mbox{ } + \alpha _2 \beta _3 [\cos (\Omega _\sigma Nt)\left| \psi \right\rangle _3- i\sin
(\Omega _\sigma Nt)\left| \psi \right\rangle _2 ]\}
\nonumber \\ &
 \mbox{ } + \exp [ - i2\Omega _s Nt]\beta _2 \beta _3 \left| \psi \right\rangle _4,
 \end{align}

\noindent
yielding the entangled state of the two nodes if only the condition \ref{cond1} is
satisfied. So here, we can vary the coupling constant $\Omega _\sigma $ and
interaction time t in some possible intervals providing a realization of
general $iSWAP$ gate with arbitrary tunable angle of rotation $\Omega _\sigma
Nt$.

Thus we can perform $\sqrt {iSWAP} $ gate which entangles the two qubits and
provides a complete set of universal quantum gates together with single
qubit operations. Here, we note that the single qubit gates can be performed
by transfer the atomic qubit to photonic qubit in waveguide where it can be
rotated on arbitrary angle by usual optical means \cite{Kok2007}. We can also return
the qubit back to QM node on demand as it has been shown above. Another
possibility to implement the single qubit gates is to transfer it to the
node with single resonant atom which state can be controlled by external
classical field. Also we can mark the principle possibility to exploit
the collective blockade mechanism for realization of the single qubit gate
similar to approach developed for usual blockade mechanism \cite{Saffman2010}.

\section{Discussion}
\label{sec:Dis}
In this work we proposed a novel architecture of solid state QC based on
multi-atomic ensembles (nodes) with integrated QM in the flexible circuit
with local QED cavities. The evaluated scheme of QC can be realized in 2D
and 3D architecture due to flexibility of the waveguide circuit providing a
quantum transport between different nodes.

For the first time we found the optimal moderate parameters of the atoms and
circuit for 100{\%} efficiency of storage and retrieval of multi qubit
photon fields that opens a door for practical realization of ideal
multi-mode quantum memory. Then we revealed  self quantum modes of the
quantum computer for reversible transfer of the qubits between the quantum
memory node and arbitrary other modes. Also, we show a realization of fast
$iSWAP$ gate for arbitrary pair of two nodes which drastically accelerates
basic quantum processes. Direct coupling of large number of the node pairs
can be simultaneously realized in the circuit that opens a practical road
for fast parallel quantum processing of many accelerated $iSWAP$ gates.

In this work, we also proposed new mechanisms for suppression of undesired
(excess) state based on the collective coherent multi-atomic dynamics and
demonstrate its application for realization of $\sqrt {iSWAP} $gate. In the
first case we use a \textit{Collective Dynamical Elimination} of the excess quantum state at some fixed moment of
times. The second collective mechanism exploiting stronger coupling with
additional local modes provides a permanent suppression of any transition to
the excess state. We note that the two proposed collective blockade mechanisms are determined by
the nonlinear collective interaction of the atoms with the electromagnetic
field modes therefore the mechanisms are free from the decoherence
in comparison with the well-known blockade based on the direct
dipole-dipole interactions.

We have demonstrated that the proposed QC architecture is scalable for
construction of many coupled distinct quantum circuits that opens a
promising way for realization of multi-qubit QC with a number of coupled
qubits limited only by the atomic coherence time. We anticipate that the
proposed atomic QC can principally work at room temperatures for example on
NV-centers in a pure diamond which looks now one of the promising candidates
for the qubit carriers up to $\sim $ $10^{ - 3}-1$ sec. timescale \cite{Balasubramanian2009}. We
also believe that promising situation for realization of $\sqrt {iSWAP} $
gate in our approach is in solid state media with atoms characterized by
large dipole moments or the quantum dots in semiconductors.

\section{Acknowledgement}
\label{sec:Ack}
The authors thank the grant of the Russian Foundation for Basic Researches
numbers: 08-07-00449, 10-02-01348 and Government contract of RosNauka {\#}
02.740.11.01.03.

\end{document}